\documentclass{myws-p10x7}

\newcommand{\x}{\mbox{$x$}}

\newcommand{\bq}{\mbox{$b$}}
\newcommand{\xb}{\mbox{$x_b$}}

\newcommand{\B}{\mbox{$B$}}
\newcommand{\xB}{\mbox{$x_B$}}
\newcommand{\xBr}{\mbox{$x^{reco}_B$}}
\newcommand{\xBm}{\mbox{$\langle x_B\rangle$}}
\newcommand{\xBLm}{\mbox{$\langle x^L_B\rangle$}}
\newcommand{\xBwm}{\mbox{$\langle x^{wd}_B\rangle$}}
\newcommand{\xbwm}{\mbox{$\langle x^{wd}_b\rangle$}}
\newcommand{\semilep}{\mbox{$D^{(*)}\ell\nu X$}}

\newcommand{\etal}{{\it et al.}}
\begin{document}

\title{
Studies of b-quark fragmentation
}

\author{V. Ciulli}

\address{
Scuola Normale Superiore, P.zza dei Cavalieri 7, 56126 Pisa, Italy
\\E-mail: Vitaliano.Ciulli@cern.ch
}

\twocolumn[\maketitle\abstract{
I will review new studies of \bq-quark fragmentation performed at the
Z peak by ALEPH and SLD. An improved sensitivity to distinguish
between fragmentation model and more accurate measurements
of the mean \bq-hadron scaled energy have been obtained.  
}]

\section{Introduction}

In $e^+e^-$ collisions the \bq-quark fragmentation function is given by 
the normalized scaled energy distribution of \bq-hadrons
\begin{equation}
D(\x)\equiv\frac{1}{\sigma}\frac{d\sigma}{d\x}
\end{equation} 
where \x\ is the ratio of the observed \bq-hadron
energy to the beam energy. 
Usually, since the \bq-quark mass is much larger than the QCD scale $\Lambda$,
the \bq-quark energy prior to hadronization is calculated using  
perturbative QCD. Hence the \bq-hadron energy is related to the quark
energy via model-dependent assumptions.   
Therefore measurement of $D(\x)$ serve to constrain both perturbative
QCD and model predictions.
Furthermore, the uncertainty in the fragmentation function $D(\x)$
must be taken into account in studies of the production and decay of
heavy quarks: more accurate measurements of this function will allow
increased precision test of heavy flavour physics.
  
At this conference new measurements have been presented by
ALEPH\cite{aleph} and by SLD\cite{sld}.  
 
\section{ALEPH measurement}

ALEPH searches for $B^+$ and $B^0$ mesons\footnote{Charge conjugation
    implied throughout} in five semi-exclusive decay
channels $\B\to\semilep$. In three of them the \B\ decays to
$D^{*+}\ell\nu X$, followed by $D^{*+} \to D^0\pi^+$, and the $D^0$
is reconstructed in the decay channels $D^0\to K^-\pi^+$, $ D^0\to
K^-\pi^+\pi^+\pi^-$ and $D^0\to K^-\pi^+\pi^0$. The remaining two
channels are $B\to D^0\ell\nu X$, followed by $D^0\to K^-\pi^+$, and 
 $B\to D^+\ell\nu X$, followed by $D^+ \to  K^-\pi^+\pi^+$.
Using the full data sample collected at the Z peak, about 4 million
hadronic Z decays, a total of 2748 candidates have been found, with a
signal purity between 63\% and 90\%, depending on the decay channel. 
The \B\ energy is estimated from $D^{(*)}$ and lepton momentum, plus the
hemisphere missing energy, due to the neutrino. The energy resolution is
described by the sum of two Gaussians, with widths of 0.04 and 0.10,
and 50-60\% of the candidates in the core. 

\begin{table}
\caption{ALEPH: results of model-dependent analysis. Errors include
systematic uncertainty}\label{tab:aleph}
\begin{tabular}{|l|c|c|} 
 
\hline 
 
\raisebox{0pt}[12pt][6pt]{Model} & 
 
\raisebox{0pt}[12pt][6pt]{\xBLm} & 
 
\raisebox{0pt}[12pt][6pt]{$\chi^2/$ndf} \\
 
\hline
 
\raisebox{0pt}[12pt][6pt]{Peterson\cite{peterson}} & 
 
\raisebox{0pt}[12pt][6pt]{$0.733\pm0.006$} & 
 
\raisebox{0pt}[12pt][6pt]{116/94} \\

\raisebox{0pt}[12pt][6pt]{Kartevelishvili\cite{kart}} & 
 
\raisebox{0pt}[12pt][6pt]{$0.746\pm0.008$} & 
 
\raisebox{0pt}[12pt][6pt]{97/94} \\
 
\raisebox{0pt}[12pt][6pt]{Collins\cite{collins}} & 

\raisebox{0pt}[12pt][6pt]{$0.712\pm0.007$} & 
 
\raisebox{0pt}[12pt][6pt]{164/94} \\\hline

\end{tabular}
\end{table}

\begin{figure}
\epsfxsize180pt
\figurebox{}{}{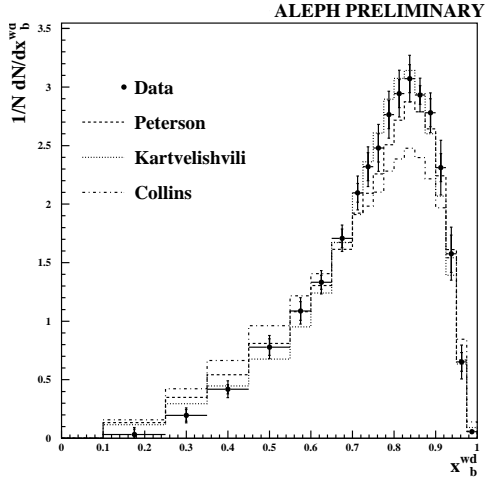}
\caption{Scaled energy distribution of weakly-decaying \B\ mesons as measured by ALEPH.}
\label{fig:aleph}
\end{figure}

The mean scaled energy \xBm\ is extracted from the raw \xB\
distribution in both a model dependent and a model-independent way. 
In the first case, three different fragmentation models have been used
to hadronize the \bq-quark after the parton shower, simulated by 
JETSET 7.4\cite{jetset}. Reconstruction efficiency and energy
resolution, as well as missing pions from $B^{**}$ and $D^{**}$
decays, are taken into account by Monte Carlo simulation.
Then, for each model, the reconstructed \xB\ spectrum in Monte Carlo is
compared to data and a minimization of the difference is performed by
varying the model parameter. Table~\ref{tab:aleph} shows the results
for \xBLm, the mean scaled energy of the leading \B, the meson
resulting from hadronization, prior to any decay. 
The fragmentation model of Kartvelishvili \etal\cite{kart} gives the
better agreement with data. 
In the model-independent analysis, the Monte Carlo is used to calculate
the
efficiency, $\epsilon(\xB)$, and the resolution matrix $G(\xB,\xBr)$,
defined as the probability for 
the \B\ meson to have a scaled energy \xB, given the measured \xBr. Hence the
fragmentation function $D(\xB)$ is obtained by unfolding the measured
distribution $D^{data}(\xBr)$: 
\begin{equation}\label{unfold}
D(\xB) = \epsilon^{-1}(\xB) \cdot G(\xB,\xBr) \cdot D^{data}(\xBr) \, .
\end{equation}
Since $G$ depends on the Monte Carlo fragmentation function, the
procedure must be iterated, using in the Monte Carlo the above $D$
function obtained from data, until convergence is reached.    
The results of this analysis are 
$
\xBLm = 0.7499\pm0.0065(stat)\pm0.0069(syst)
$
, for the leading \B\ meson, and
$
\xBwm = 0.7304\pm0.0062(stat)\pm0.0058(syst)
$
, for the weakly decaying one.
Figure~\ref{fig:aleph} shows the resulting fragmentation function for
the weakly-decaying B meson, compared to the distributions obtained
from the model-dependent analysis. 

\section{SLD measurement}
      
SLD measurement is based on 350,000 $Z$ hadronic decays collected in
97 and 98. The analysis method is the same used for an already published
SLD
result\cite{oldsld}, based on a smaller data sample. 
A topological secondary vertex finder exploits the small and stable SLC
beam spot and the CCD-based vertex detector to inclusively reconstruct
\bq-decay vertices with high efficiency and purity. Precise vertexing
allows to reconstruct accurately the \bq-hadron flight direction and
hence the transverse momentum of tracks associated to the vertex with
respect to this direction. Their invariant mass, corrected for the
transverse momentum of missing particle, is used to separate
\bq-hadrons from $udsc$ background, yielding a 98\% pure \bq-sample
with 44\% efficiency.   
 
The \bq-hadron energy is also measured from the invariant mass and the
transverse momentum of the tracks associated to the
vertex. Constraining the vertex mass to the $B^0$ mass, an upper limit
on the mass of the missing particles is found for each reconstructed
\bq-decay vertex, and is used to solve for the longitudinal momentum
of the missing particles, and hence for the energy of the \bq-hadron.  
In order to further improve the \bq-sample purity and the
reconstructed \bq-hadron energy, only vertices with low invariant mass
are kept. The selection yields 4164 candidates, with an overall
efficiency of 4.2\% and 9.6\% energy resolution in the core, which
accommodate about 80\% of the candidates. Moreover, both the efficiency
and the energy resolution are remarkably flat in the region $\xb > 0.2$   
 
%

\begin{figure}
\epsfxsize180pt
\figurebox{}{}{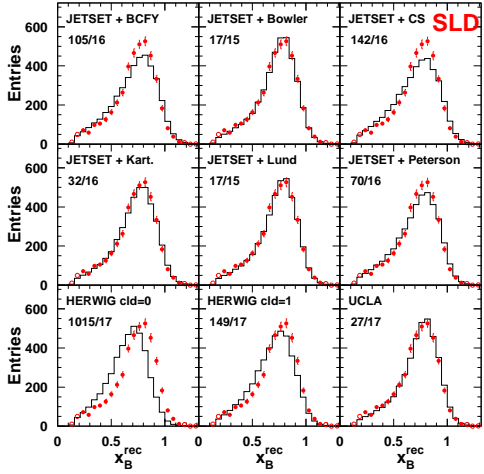}
\caption{SLD test of fragmentation models.}
\label{fig:sld}
\end{figure}

Several fragmentation functions have been fitted to the candidates
scaled energy distribution, as shown in Figure~\ref{fig:sld}. 
A good description of the data is obtained using JETSET 
together with the phenomenological models of the Lund
group\cite{lund}, Bowler\cite{bowler} and Kartvelishvili {\em et
al.}\cite{kart}, or using the UCLA\cite{ucla} fragmentation model.
Several functional forms of the true
energy distribution $D(\xb)$ have been tried too, and four of them
have been found consistent with data. 
Hence the true distribution has been obtained from
equation~\ref{unfold}, using the above eight best fitted distributions
to calculate the unfolding matrix $G$ from Monte Carlo. 
The resulting mean scaled energy for the weakly-decaying \bq-hadron is 
$
\xbwm = 0.710\pm0.003(stat)\pm0.005(syst)\pm0.004(model) 
$.
\begin{figure}
\epsfxsize180pt
\figurebox{}{}{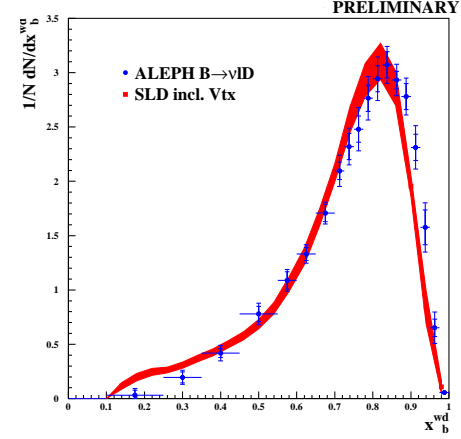}
\caption{Scaled energy of the weakly-decaying \bq-hadron as measured
by ALEPH and SLD.}
\label{fig:comp}
\end{figure}

\section{Summary and conclusions}

In Figure~\ref{fig:comp}, the fragmentation functions measured by
ALEPH and SLD are compared. A slight disagreement is observed between
the two. However it must be pointed out that ALEPH selects 
\B\ mesons only, while the SLD sample also includes $B_s$ and barions, which 
may be responsible of the observed difference.

\end{document}